# Deep learning methods based on cross-section images for predicting effective thermal conductivity of composites

Qingyuan Rong [a], Han Wei [a], Hua Bao [a, *]

[a]*University of Michigan—Shanghai Jiao Tong University Joint Institute, Shanghai Jiao Tong University, Shanghai 200240, China*

Effective thermal conductivity is an important property of composites for different thermal management applications. Although physics-based methods, such as effective medium theory and solving partial differential equation, dominate the relevant research, there is significant interest to establish the structure-property linkage through the machine learning method. The performance of general machine learning methods is highly dependent on features selected to represent the microstructures. 3D convolutional neural networks (CNNs) can directly extract geometric features of composites, which have been demonstrated to establish structure-property linkages with high accuracy. However, to obtain the 3D microstructure in composite is generally challenging in reality. In this work, we attempt to use 2D cross-section images which can be easier to obtain in real applications as input of 2D CNNs to predict effective thermal conductivity of 3D composites. The results show that by using multiple cross-section images along or perpendicular to the preferred directionality of the fillers, the prediction accuracy of 2D CNNs can be as good as 3D CNNs. Such a result is demonstrated with the particle filled composite and a stochastic complex composite. The prediction accuracy is dependent on the representativeness of cross-section images used. Multiple cross-section images can fully determine the shape and distribution of fillers. The average over multiple images and the use of large-size images can reduce the uncertainty and increase the prediction accuracy. Besides, since cross-section images along the heat flow direction can distinguish between serial structures and parallel structures, they are more representative than cross-section images perpendicular to the heat flow direction.

**1. Introduction**

    Composites have wide applications in electronics, aerospace and renewable energy systems [1, 2]. For example, carbon nanotube-polymer composites are promising thermal interface materials for electronics cooling [3, 4], benefiting from the extremely high thermal conductivity of carbon nanotubes (CNTs) [5]. The lightweight CNT-polymer composites are also best

---

* Corresponding author. University of Michigan—Shanghai Jiao Tong University Joint Institute, Shanghai Jiao Tong University, Shanghai 200240, China.
  E-mail address: hua.bao@sjtu.edu.cn.

suited for the highly demanding heat dissipation requirements of spacecraft components [6-8]. Besides, polypyrrole-silver (PPy-Ag) composites films have the potential for thermoelectric power generation [9], because the suppression of thermal conductivity can improve the efficiency defined by figure-of-merit [10]. Effective thermal conductivities of composites can largely influence their performance. They are also needed for high-fidelity modeling of physical processes like combustion of propellants [11]. Meanwhile, efficiently predicting the effective thermal conductivity of composites can help establish structure-property linkages for accelerated materials discovery and design [12]. However, since the effective thermal conductivity of a composite is a complex function of the thermal conductivity of the constituents, the filler volume fraction, the filler shape and size, the filler dispersion, and the thermal interfacial resistance, the prediction remains as a critical task [2, 13].

Numerous studies have been carried out to predict the effective thermal conductivities of composites, including theoretical models based on heat-transfer mechanisms and simulation models based on accurate characteristics of microstructures [1]. In theoretical models, heterogenic materials are considered as being macroscopically homogenic, which is known as effective medium theory (EMT), or effective medium analysis (EMA) [14]. The fundamental model of the EMT is the Maxwell model [15] proposed more than a hundred years ago, which is suitable for lower volume fraction of particle fillers because particles are assumed to be far apart from each other. After that, lots of improvements have been made. For example, Hasselman *et al*. [16] modified the original Maxwell model to include the influence of interfacial thermal resistance. Xu *et al*. [17] proposed a reconstruction of Maxwell model for composites consisting of continuum matrix and particle fillers. At higher volume fraction, extensive information like correlations between the positions of particles is required. Bruggeman model [18, 19] offers a simple method to estimate these interaction effects from the premise that the fields of neighboring particles can be taken into account by adding the dispersed particles incrementally. Every *et al*. [20] presented a phenomenological model in which particle size is considered as a factor that can influence the effect of interfacial thermal resistance. As for particle shape, Nan *et al*. [21] proposed general EMA formulations considering ellipsoidal particle fillers, where the shape of a particle is described using a simple aspect ratio. Despite these progresses, EMTs usually have low prediction accuracy for composites close to and beyond the percolation threshold [14]. They are also difficult to be generally applied to anisotropic composites with complex microstructures[1]. As for simulation models, molecular dynamics [22], lattice Boltzmann method [23] and finite element method (FEM) [24-26] are commonly used in microscale, mesoscale and macroscale, respectively. Their common drawbacks include the difficulty in generating complex microstructures, and the requirement of high computing powers and time [14].

In recent years, data-driven approaches have shown enormous promise within materials science [27-29], and many have achieved practically useful combinations of high prediction accuracy and low computational cost [30, 31]. For example, Gupta *et al.* [32] use principal components analysis to select microstructure features from *n*-point special correlations, which are further used to extract robust structure-property linkage for complex non-metallic inclusion/steel composite system. Liu *et al.* [33] develop multi-agent learning framework which combines high-level features based on either geometric descriptors or physics-related functions and microscale features based on voxel neighborhood characteristics [34] separately. Such strategy improves prediction accuracy for complex hierarchical material systems. One of the most important factors for these machine learning approaches is the features used, which can largely influence the prediction accuracy [35]. Although *n*-point spatial correlations provide a naturally organized set of measures of the microstructure, the complete set is extremely large and unwieldy even for $n = 2$. Higher-order spatial correlations (i.e., 3-point and higher) have been out of practical reach so far, largely due to the unwieldy size of the complete space of such descriptors [32]. On the other hand, the emergence of deep learning methods based on convolutional neural networks (CNNs) allows automatic discovery of features needed from extremely large set of potential features [36, 37]. Since the explicit feature design is not required, the trained models usually exhibit higher generalization properties [38]. In early 2016, Liu *et al.* [39] presented a large-scale, big data application of deep learning in materials discovery. The deep convolutional neural networks based on experimental images achieved state-of-the-art accuracy in an electron imaging indexing problem. In 2018, our previous work [40] demonstrated that deep learning methods can be used as a fast prediction tool to obtain the effective thermal conductivity of 2D isotropic composites with high accuracy. As for 3D composites, Yang *et al.* [41] predicted the effective stiffness of anisotropic elastic composites using 3D CNNs, which outperforms the sophisticated physics-inspired approaches by 54%. Cecen *et al.* [42] formulated an integrated framework combining 3D CNN features as estimators of higher order statistics with 2-point spatial correlations, which shows improved accuracy in predicting effective properties of 3D composites. They also claim that the missing information of 3D CNN features may eventually be learned by a more complicated and heavily layered network, or with the inclusion of more data. Although the prediction accuracies of 3D CNNs trained based on 3D microstructures of composites can be high, it can also be difficult to obtain the detailed 3D microstructures in real applications. For example, 2D microscopic images of composites are usually much easier to get in experiments. Therefore, predicting effective thermal conductivity of 3D composites from 2D cross-section images can be very useful.

There are several different methods for 3D reconstruction tasks investigated in computer vision area. For example, the algorithm of matching cubes can generate 3D tissue or bone surfaces from multiple 2D slices of medical data [43]. 3D reconstructions of buildings or humans can also be achieved from several images taken in different directions [44], single still image whose depth is modeled by Markov Random Field trained via supervised learning [45], or video sequences [46]. In this work, we aim to predict effective thermal conductivities of 3D composites based on 2D cross-section images. We will show that deep learning methods can automatically infer 3D structure properties from multiple cross-section images in different directions. Due to the lack of a suitable experimental data, we assume that the ground truth is reasonably well captured by the results of finite element models applied on digitally generated microstructures. Two different microstructure generation methods are used to demonstrate the generalization property on different types of microstructures, including particle filled composites and stochastic complex composites. Optimizations for CNN architectures, like deeper network, batch normalization and dropout, are applied to improve the prediction accuracy. The influences of the dataset size, and the size, direction and number of cross-section images are also investigated. The results of 2D CNNs are further compared with 3D CNNs and theoretical models. Finally, we will discuss the representativeness of cross-section images for revealing 3D structure information.

## 2. Methods

*2.1 Dataset generation*

We use two different methods to generate microstructures of 3D composites, which are packing method and QSGS method for composites with spherical particles and composites with stochastic complex geometries respectively. The generated structures are then used to calculate effective thermal conductivities by FEM. The ratio of sample numbers for training, validating and testing is 4: 1: 1, where maximum number of training samples reaches up to 2000.

Packing method is an efficient numerical tool to generate systems with hard spherical particles [47, 48]. One of the fundamental packing algorithms is called Lubachevsky–Stillinger (LS) algorithm [49] developed in 1990, which processes a system of randomly moving spherical particles whose size grows in time. Since the particle size is increasing during collision, an extra velocity is added after collision to avoid overlapping of two spheres. The velocities of two spheres after collision can be calculated by following equations:

$$\mathbf{v}_1^* = \left[\mathbf{v}_2^{(p)} + h\mathbf{u}_{12}\right] + \mathbf{v}_1^{(t)} \tag{1}$$

$$\mathbf{v}_2^* = \left[\mathbf{v}_1^{(p)} + h\mathbf{u}_{21}\right] + \mathbf{v}_2^{(t)} \tag{2}$$

where *p* represents parallel component, *t* represents transverse component, *h* is set as the diameter growth rate and **u** is the unit vector between two particles with index 1 and 2.

In this work, LS algorithm is implemented using event-driven molecular dynamics [50], which predicts the nearest collision between all spheres and then update their positions and velocities. Periodic boundary conditions are applied to a unit cubic cell and cell method is used to improve the computational efficiency. For all dataset samples, the number of spheres is fixed as 100 and the volume fraction of spheres ranges from 0.1 to 0.6. The generated mono-disperse sphere packing systems are further uniformly discretized using a mesh of 100×100×100 after convergence test, so the diameter of spheres ranges from about 12.6 to 22.6 in terms of mesh size. Due to the nature of LS algorithm and the spherical shape, the generated composites are isotropic. Three examples with volume fraction of 0.1, 0.35 and 0.6 are shown by Fig. 1(a)~(c).

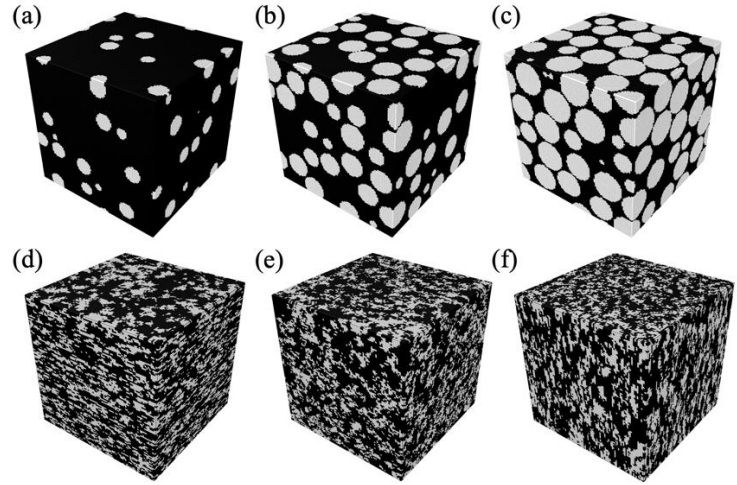

FIG. 1. Microstructures of composites with size (voxel number) of 100×100×100. (a)~(c) Composites with spherical particles generated by packing method, whose volume fractions are 0.1, 0.35 and 0.6 respectively. (d)~(f) Composites with stochastic complex geometries with different directionalities generated by QSGS method, whose volume fractions are fixed as 0.35.

Quartet structure generation set (QSGS) is developed by Wang *et al*. [51] as a comprehensive approach in which four parameters are identified for controlling the microstructures of porous media. The advantage of this method is that the morphological features formed are closely resembling the forming progress of many real porous media. There are basically two steps in QSGS method. The first step is to randomly generate cores in the system based on a parameter called core distribution probability, which should not be larger than the final volume fraction. The second step is to randomly grow from these cores, until the required volume fraction is reached. The growing probabilities in different directions can be set as different values, which causes the final structures to be anisotropic.

We also extend the aforementioned QSGS method to 3D space. The principles are the same as the 2D case, but the number of possible growth directions of a core increases from 7 (= $2^3$-1) to 26 (= $2^3$-1). By controlling these parameters, we can generate anisotropic composites. In this work, we generate anisotropic composites with preferred directionalities in the three perpendicular directions, as shown in Fig. 1(d)~(f). Since we focus on the effect of anisotropy, the volume fraction is fixed as 0.35 for all structures. Two different sizes, i.e., 100×100×100 and 50×50×50, are used to investigate the influence of image size on the prediction accuracy. The QSGS parameters for different sizes are set as the same, which means the small-size composites can be viewed as a small part of large-size composites.

We use a python package called fenics developed by the FEniCS Project [52] to solve the heat diffusion equation for 3D composites and obtain the effective thermal conductivities. The thermal conductivities of the filler and the matrix are set as 10 and 1, respectively. The typical temperature distribution at steady state is shown in Fig. 2(a). The temperatures in the top surface and the bottom surface are fixed as 1 and 0, respectively. Periodic boundary conditions are used for the rest of the surfaces. The heat transfer rate along $z$ axis equals to the extracted energy from the top surface or the injected energy into the bottom surface, which is used to calculate the effective thermal conductivity.

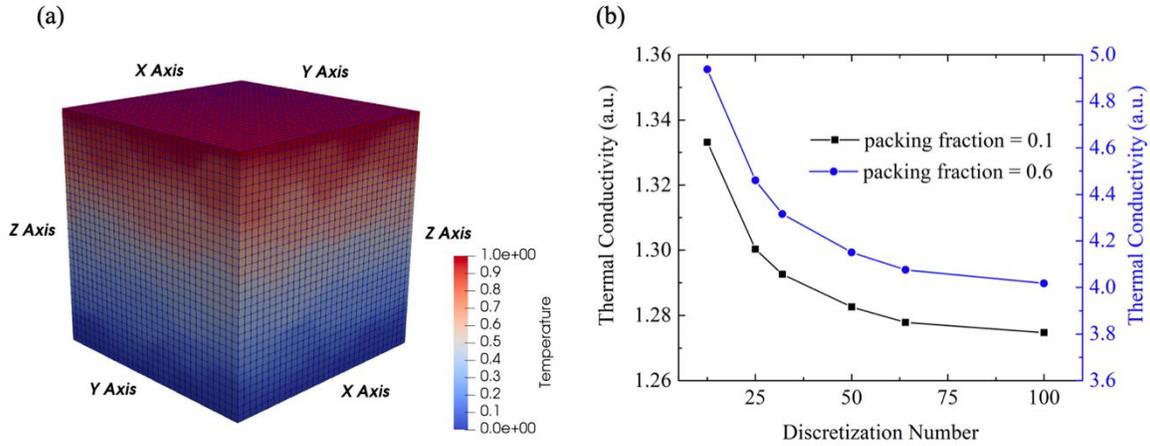

FIG. 2. (a) Steady-state temperature distribution calculated by FEM. (b) Convergence test for discretization of packing systems with spherical particles. The result almost converges with discretization number of 100.

In order to get accurate results from FEM, convergence tests are necessary to choose a set of reasonable parameters. We first test the discretization of each voxel in microstructures generated from packing method or QSGS method, which converges when each voxel is further divided by 8 (= $2^3$) small voxels for both hexahedron and tetrahedron cell types. We choose to use hexahedron cell type because the computational cost is smaller. We further test the discretization for packing method, since the spherical fillers will inevitably cause discretization errors for the hexahedron cell type. The results are shown in Fig. 2(b). For both packing fractions of 0.1 and 0.6, the effective thermal conductivity converges as the discretization number increases to 100. Therefore, for isotropic composites generated by packing method, total mesh number for final calculations is 200×200×200, and for anisotropic composites generated by QSGS method, total mesh numbers are 200×200×200 and 100×100×100 for large-size structure (100×100×100) and small-size structure (50×50×50), respectively.

*2.2 Deep learning methods*

Deep learning methods are representation-learning methods with multiple levels of representation, which means they allows a machine to be fed with raw data and to automatically discover the representations needed [36]. Typically, deep leaning methods are used for detection or classification problems. Since the output in this work is the effective thermal conductivity, i.e. a scaler, deep learning methods are actually used to train regression models. Convolutional neural networks are one particular type of deep learning method that are designed to process multiple arrays of data, such as 2D images and 3D videos. Typically, CNN is structured as a series of stages, where the first few stages are composed of convolutional layers to detect local conjunctions of features and pooling layers to merge semantically similar features into one. By composing non-linear modules that each transform the representation at one level (starting with the raw input) into a representation at a higher, slightly more abstract level, very complex functions can be learned.

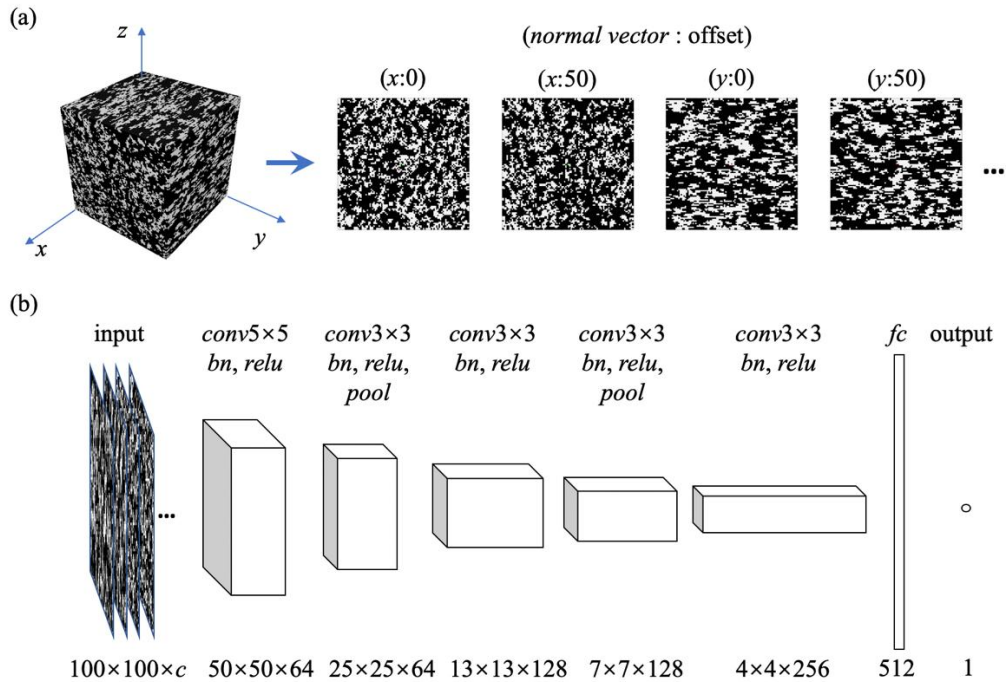

FIG. 3. (a) Multiple cross-section images from one 3D structure. The index of an image is formed by two components, i.e. normal vector of the image and the offset in that direction. (b) Typical CNN architecture used in this work, where *conv* followed by the kernel size represents convolutional layer, *bn* represents batch normalization layer, *relu* represents rectified linear unit, *pool* represents max pooling layer and *fc* represents fully connected layer. The input is composed of $c$ cross-section images and the output is the effective thermal conductivity.

A typical deep learning model based on cross-section images is shown in Fig. 3. The first step is to obtain a series of cross-section images from a 3D structure, as shown in Fig. 3(a). The index of a cross-section image is denoted by a pair of the normal vector and the offset. The cross-section images are simply stacked in certain order to serve as different channels of the input image for the CNN, as shown in Fig. 3(b), where the number of cross-section images is denoted by $c$. The effects of number and direction of cross-section images will be investigated. The influence of different network architectures will also be investigated, including the network depth, the use of batch normalization [53] and the use of dropout strategy [54]. The loss function used to train the models are chosen as root mean square error loss:

$$loss = \sqrt{\frac{1}{N}\sum_{i=1}^{N}(y_i - \hat{y}_i)^2} \qquad (3)$$

where $y_i$ is the target value from FEM and $\hat{y}_i$ is the predicted value from CNN model. Adam algorithm is chosen to perform the backpropagation [55] that updates the parameters of CNN models. After training, the models are saved and tested on a separate dataset. Two different criteria are used to quantify the prediction accuracy, i.e. mean absolute error (MAE) and root mean square error (RMSE) defined by following equations:

$$\text{MAE} = \frac{1}{N}\sum_{i=1}^{N}\left|\frac{y_i - \hat{y}_i}{\bar{y}}\right| \times 100\% \qquad (4)$$

$$\text{RMSE} = \sqrt{\frac{1}{N}\sum_{i=1}^{N}\left(\frac{y_i - \hat{y}_i}{\bar{y}}\right)^2} \times 100\% \qquad (5)$$

where $y_i$ is the target value from FEM, $\hat{y}_i$ is the predicted value from CNN model, and $\bar{y}$ is the average value of the testing dataset samples. MAE quantifies the absolute prediction error normalized by the mean of the prediction value and RMSE quantifies the root square error of prediction normalized by the mean. Although the values of the two criteria are slightly different, they have similar meanings and both of them decrease with the loss defined by Eq. 3.

The results of 2D CNNs explained above will be compared with 3D CNNs where the input is directly the 3D structure. The network architecture of 3D CNNs are very close to 2D CNNs, except that the convolutional kernels for 3D CNNs are extended from 2D to 3D, for example, from 5×5 to 5×5×5. Consequently, the convolutional kernels can slide in all the three directions to capture 3D local structure information. In comparison, for single-channel or multi-channel 2D CNNs, the convolutional kernels can only slide in the image plane, which means the structure information in the third direction must be inferred by single or multiple cross-section images.

## 3. Results

In order to improve the prediction accuracy of CNNs, the recent trend has been to increase the number of layers [56], while using dropout [54] to address the problem of overfitting. The deep CNNs require a large dataset to train and batch normalization [53] is a very successful strategy to accelerate training process and alleviate overfitting. Therefore, based on particle filled composites that are relatively simple, we first carry out a series of optimizations, including the depth of CNNs, the use of batch normalization, the number of cross-section images, the dataset size and the use of dropout. Then, we apply the optimized model to train on stochastic complex composites to demonstrate its generalization property. Finally, the representativeness of cross-section images will be discussed.

*3.1 Performance on particle filled composites*

The design of the CNN architectures is based on AlexNet [57], which competed in the ImageNet Large Scale Visual Recognition Challenge in 2012. Two different depths are tested, i.e., CNNs with 3 and 5 convolutional layers. Batch normalization [53] is further applied to the deep CNN. Usually, no depth can be computed from a single image without priori information [44]. Therefore, multiple cross-section images are used in order to improve the prediction accuracy. Specifically, four cross-section images, i.e., "*x*:0", "*x*:50", "*y*:0" and "*y*:50", are used to investigate the influence of CNN architectures. The reason that we do not choose *z* direction images is that microstructures along the heat flow direction can have larger influence on the effective thermal conductivity, which will be discussed latter. The sizes of training dataset, validating dataset and testing dataset are 800, 200 and 200, respectively. As shown in Fig. 4(a), performance of network with 5 convolutional layers is better than network with 3 convolutional layers. This is because deeper networks can capture more abstract features of the input images. Besides, the performance of the network with batch normalization is much better, which demonstrates that batch normalization can alleviating overfitting problems. In the following work, we will use network with 5 convolutional layers and batch normalization, i.e., architecture shown in Fig. 3(b).

We further investigate the performance of CNNs with different numbers of cross-section images. The results are shown in Fig. 4(b). For 1 image, since the microstructures of composites are isotropic, we simply choose "*x*:0". For 2 to 8 images, we equally choose images in *x* and *y* directions that are uniformly distributed. As the results show, testing error decreases with number of images. This is because more images together can represent more information about the structure, especially for properties like volume fraction that relies on the whole of the structure.

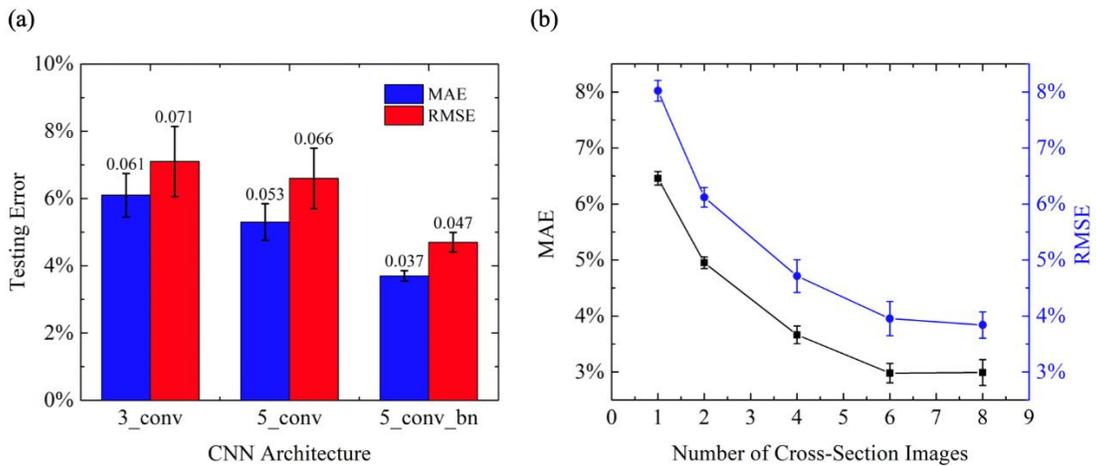

FIG. 4. Testing error of 2D CNNs when sizes of datasets for training, validating and testing are 800, 200 and 200. (a) Dependence on different CNN architectures. CNNs with 3 and 5 convolutional layers are denoted by 3_conv and 5_conv, respectively. Batch normalization is further added to 5_conv, which is denoted by 5_conv_bn. Four cross-section images, i.e., "*x*:0", "*x*: 50", "*y*:0" and "*y*:50" are used as input. Deep CNNs with batch normalization shows better performance. (b) Dependence on the number of cross-section images. CNN architecture with 5

convolutional layers and batch normalization (5_conv_bn) is used. Multiple images are equally chosen in *x* and *y* directions that are uniformly distributed. Testing error decreases with the number of cross-section images.

The training dataset size can have large influence on the performance of CNNs, since a small dataset can easily cause the model to overfit the training samples. The results of dataset size dependence are shown in Fig. 5(a). As the results show, both MAE and RMSE decrease as the dataset size increases. The decrease trend slows down as the size further increases.

The last strategy used in this work to avoid overfitting is the dropout method, where keep probability controls the dropout extent [54]. Keep probability of 1 means no dropout is used. As shown in Fig. 5(b), dropout influences little on testing error. This may due to the use of batch normalization that already alleviates overfitting to some extent. Meanwhile, the use of dropout increases the prediction uncertainty. The final choice of keep probability is 0.9, which shows slightly better performance.

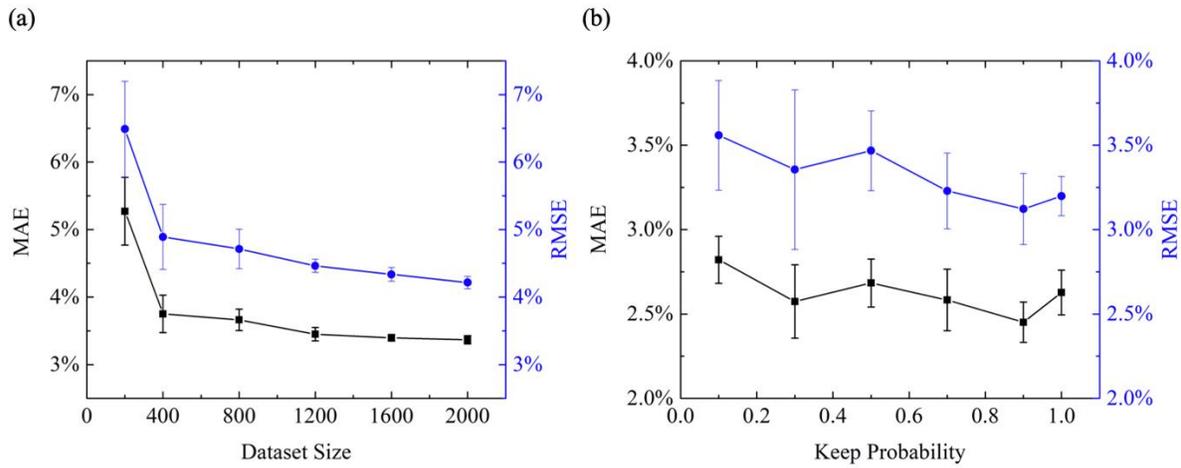

FIG. 5. Testing error of 2D CNNs with optimized architecture. (a) Dependence on dataset size. Four cross-section images, i.e., "*x*:0", "*x*: 50", "*y*:0" and "*y*:50" are used as input. Testing error decreases with the dataset size. (b) Dependence on keep probability of dropout method. Eight cross-section images are used. Sizes of datasets for training, validating and testing are 2000, 500 and 500. The dropout has small influence on testing error.

After the optimizations above, CNN architecture with 5 convolutional layers, batch normalization and dropout with keep probability of 0.9 is chosen. When training dataset contains 2000 samples and the input of the CNN model is 8 cross-section images, the lowest testing error obtained is 2.5% in terms of MAE and 3.1% in terms of RMSE, as shown in Fig. 6. The training process takes 4 hours on a single CPU, which means iteration over a sample takes about 0.015 second.

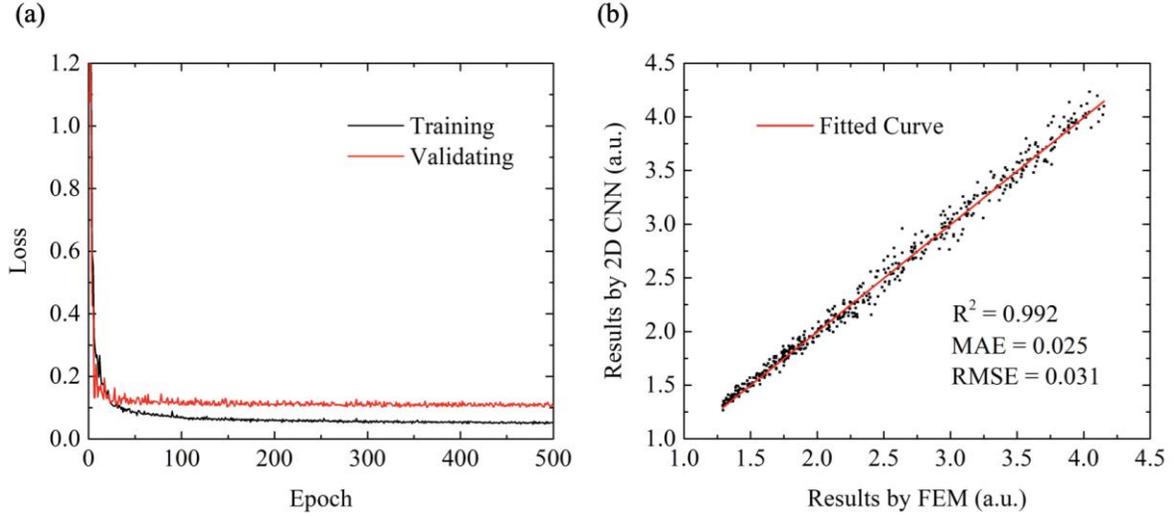

FIG. 6. (a) Training loss and validating loss during training process. (b) Comparison between the effective thermal conductivity calculated by FEM and 2D CNN model. 5 convolutional layers, batch normalization, dropout with keep probability of 0.9, eight cross-section images and training dataset size of 2000.

Next, we compare the results of 2D CNNs with 3D CNNs and analytical models. The results are shown in Fig. 7. Since 3D CNNs contains much more parameters, the training process is extremely slow, with the iteration time over a sample of about 3.3 second. For simplicity, the 3D CNNs only contains 3 convolutional layers and only 800 training samples are used, which already costs 6 days on a single CPU to train. As shown in Fig. 7(a), the obtained MAE is 3.3% and RMSE is 3.9%. The testing error of 3D CNNs is close to 2D CNNs using 6 cross-section images and the same dataset size, which demonstrates that 2D CNNs based on cross-section images can be used to infer 3D structure information. Fig. 7(b) shows the effective thermal conductivity calculated by FEM and three different effective media theories changing with packing fraction. The results show that effective thermal conductivities increase with packing fraction of spheres, since the thermal conductivity of spherical fillers is larger than the matrix. The prediction errors of Maxwell model [20] and reconstruction Maxwell model [17] are relatively large when the packing fraction is large. It is because details of microstructures are ignored in these models. However, results of Bruggeman model [19] match very well with FEM because the model can be derived based on the system of sphere packing, which is consistent with previous work [58].

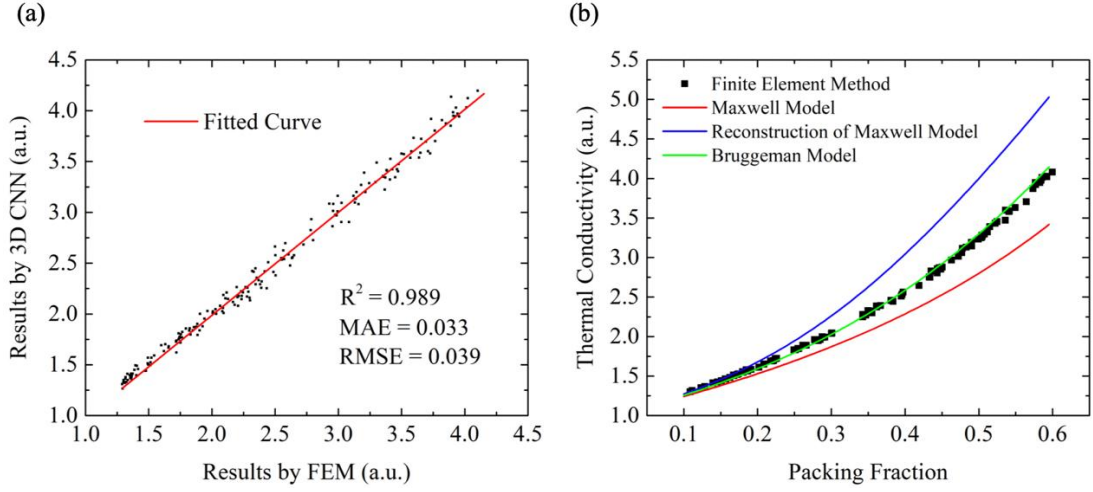

FIG. 7. Comparison between the effective thermal conductivity of particle filled composites calculated by FEM and (a) 3D CNN model, (b) effective media theories.

Although the accuracy of CNN models is not as good as Bruggeman model based on volume fraction, the prior knowledge that effective thermal conductivity for sphere packing system highly depends on packing fraction is not needed for CNN models. What's more, the CNN models even do not know the physical quantity of volume fraction. Since CNNs are able to extract features from dataset automatically, they can be easily generalized to different types of structures. We will further demonstrate this by composites with stochastic complex geometries.

*3.2 Performance on stochastic complex composites*

The comparison between FEM and aforementioned theoretical models is shown in Fig. 8(a). The volume fraction of stochastic complex composites is fixed as 0.35, and the effective thermal conductivity calculated by FEM ranges from about 2.3 to 3.3 because of the anisotropic microstructures. Since the volume fraction is a constant, the predictions of Maxwell model, reconstructed Maxwell model and Bruggeman model will be constants too, which means their prediction errors are actually related to the variation of the target values. Specifically, the testing errors of these theoretical models are 28.4%, 12.7%, 20.7% in terms of MAE and 30.4%, 14.1%, 23.4% in terms of RMSE, respectively. In contrast, using exactly the same 2D CNN model optimized for particle filled composites but trained on the dataset with 2000 stochastic complex composites, the obtained MAE is only 0.8% and RMSE is only 1.0%, which are slightly better than 3D CNNs trained on 800 stochastic complex composites, i.e. 0.9% and 1.2%, respectively. Due to the extremely large computational cost of training 3D CNNs, it is difficulty to further increase the size of training dataset. Note that the values of MAE and RMSE are somehow related to the distribution of the target values. If the differences between target values are small, then predictions from CNN models will also be within that small range, which means even random predictions can have small MAE or RMSE. This is exactly the case for the difference between the predictions of particle filled composites and stochastic complex composites. For the dataset of particle filled composites, the range of target values is about 1.1 times of the mean value, while for dataset of stochastic complex composites, the span is only about 36% of the mean value. Taking this into consideration, we further normalize the MAE and RMSE with the normalized standard deviation of corresponding target values, i.e., 33.0% and 10.9% for particle filled composites and stochastic complex composites, respectively. Hence, for particle filled composites, the normalized MAE and RMSE are 7.6% and 10.2% respectively. For stochastic complex composites, the normalized MAE and RMSE are 7.3% and 9.2% respectively, which are close to particle filled composites. The similarity demonstrates the good generalization property of CNN models.

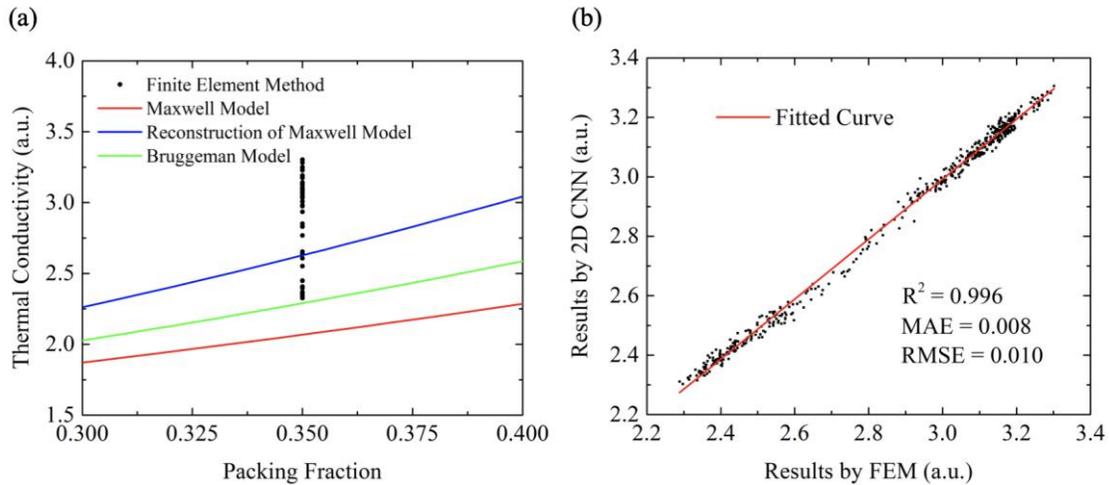

FIG. 8. Comparison between the effective thermal conductivity of stochastic complex composites with 100×100×100 size calculated by FEM and (a) effective media theories, b) 2D CNN with 5 convolutional layers, batch normalization, dropout with keep probability of 0.9, eight cross-section images and training dataset size of 2000.

We also investigate the dependence on the number of cross-section images. As shown in Fig. 9(a), for one image case, the testing error drops about 35% when the direction of cross-section image changes from perpendicular to heat flow to along heat flow, which means cross-section images along heat flow are more representative than cross-section images perpendicular to heat flow. This is because the images along the heat flow directions can directly distinguish between serial structures or parallel structures, while images perpendicular to the heat flow are not able to. By using two perpendicular images along the heat flow, the prediction error further decreases about 25%, because such combination can actually determine the shape and distribution of the fillers. Further increasing the number of cross-section images does not change the error much. This is because the property of anisotropy is not like volume fraction that requires the average over multiply cross-section images.

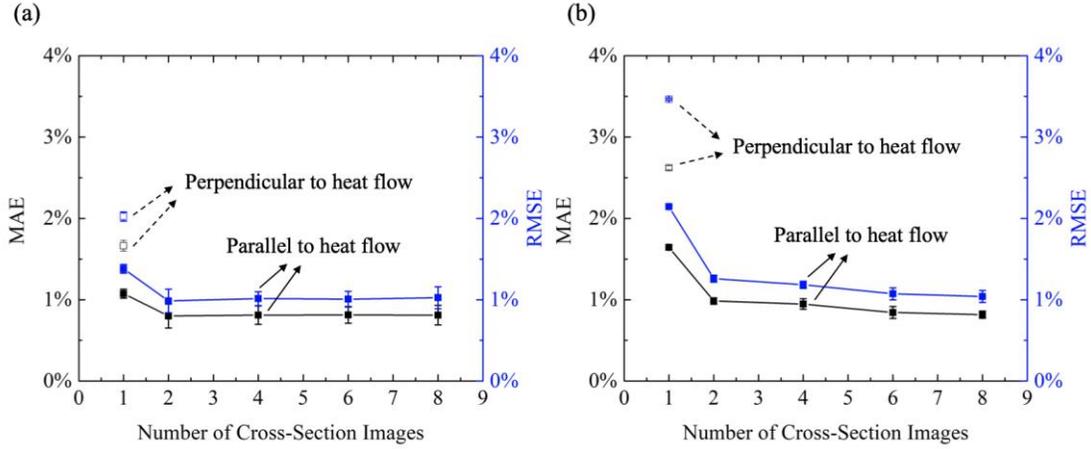

FIG. 9. Testing errors of stochastic complex composites changing with the number and direction of cross-section images. For the solid dots, cross-section images parallel to the heat flow direction, i.e., with $x$ or $y$ normal vector, are used. For the hollow dots, cross-section images perpendicular to the heat flow, i.e., with $z$ normal vector, are used. Multiple images are equally chosen in $x$ and $y$ directions that are uniformly distributed. (a) Composites with $100 \times 100 \times 100$ size. (b) Composites with $50 \times 50 \times 50$ size. Testing errors are smaller when cross-section images parallel to the heat flow direction, multiple cross-section images or cross-section images with larger size are used.

Since the size of cross-section images can also be an important factor that influences the results, we further decrease the size of the structures to $50 \times 50 \times 50$. Using similar network with eight cross-section images, but decreasing the size of input images from $100 \times 100$ to $50 \times 50$, and removing one pooling layer to keep the same depth of CNN model, the testing results are shown in Fig. 9(b). The testing error also decreases as the number of cross-section images increases and there is also a drop between 1 and 2 cross-section images. However, if only one cross-section image is used, the error is 50% larger than large-size images. If two cross-section images are used, the difference of testing error decreases to about 25%. The testing error slightly decreases when the number of cross-section images increases to eight, which demonstrates that small-size images are less representative than large-size images. For eight cross-section images, the obtained MAE and RMSE are also 0.8% and 1.0% respectively, which are the same as large-size images. We will further discuss the representativeness of cross-section images in the following chapter.

### 4. Representative cross-section images

The prediction accuracy of 2D CNN models is dependent on the representativeness of input cross-section images. If cross-section images are not able to fully represent the 3D information needed for predicting the effective thermal conductivity, the loss of information will certainly cause the decrease of prediction accuracy. Hence, we further demonstrate how to choose representative cross-section images.

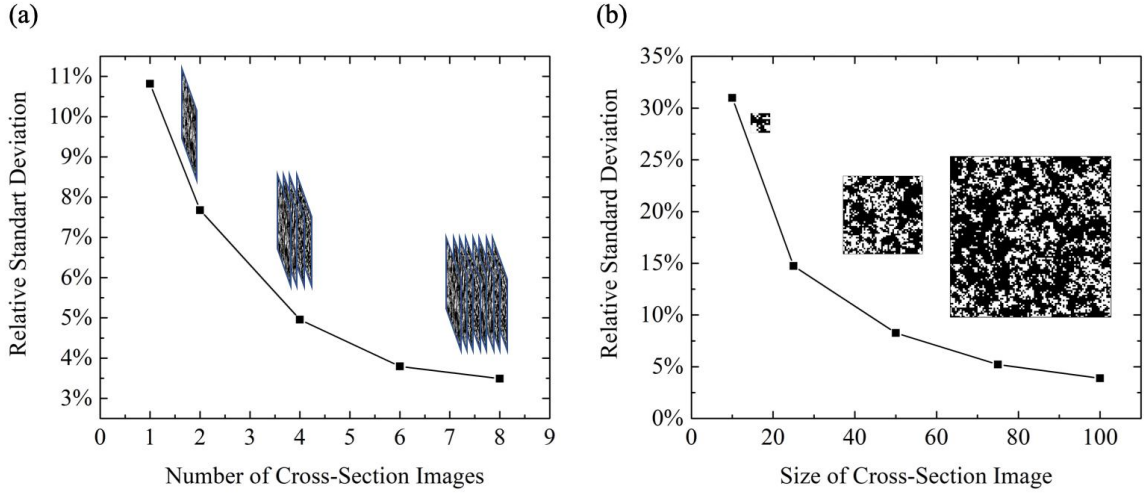

FIG. 10. (a) Relative standard deviation of cross-section images' volume fraction changing with the number of cross-section images. The size of images is 100×100, and the volume fraction of 3D structure is 0.35. (b) Relative standard deviation of cross-section images' volume fraction changing with the size of cross-section images. The volume fraction of 3D structure is 0.35.

We use the volume fraction of the cross-section images as a criterion for representing 3D information. As shown by Fig. 10(a), the standard deviation of cross-section images' volume fraction decreases with the number of cross-section images, similar to the decrease trend of testing errors of 2D CNNs shown in Fig. 4(b). This is because by taking cross-section images in different positions and then taking their average, the overall uncertainty can be reduced. Hence, multiple cross-section images together are more representative than a single cross-section image with the same size. Besides, the size of the cross-section images can also influence their representativeness. As shown in Fig. 10(b), the standard deviation of cross-section images' volume fraction decreases as the image size increases, which means larger images can be more representative than smaller ones. This is because larger images can contain more information of the structure, which reduces the chance of missing the major features of the structure. The results are also consistent with the testing error of 2D CNNs based on cross-section images with different sizes.

On the other hand, the representativeness of cross-section images is also dependent on their directions. For isotropic materials, theoretically, one cross-section image that is large enough can fully represent the 3D structure. But for anisotropic materials, one cross-section image is usually insufficient to fully capture the 3D microstructure, since it is difficult to known how fillers grow in the direction perpendicular to the cross section. However, based on cross sections along and perpendicular to the preferred directions in which the fillers grow, the 3D structure information can still be inferred. Besides, in terms of thermal conductivity, the direction of heat flow also matters. Since cross-section images along the heat flow direction can directly distinguish parallel structure or serial structure, they are more representative than images perpendicular to heat flow.

## 5. Discussions and Summary

*5.1 Discussions*

Currently, general machine learning methods usually require geometric descriptors or physics-related features in establishing structure-property linkage of composites. One of the limitations is that their prediction accuracy is highly dependent on the features selected from the extremely large space of descriptors. Since CNNs can automatically and efficiently extract useful features from the microstructures, there is no need of feature engineering, which also results in their good generalization ability to different kinds of structures.

Compared with 3D CNNs directly based on 3D microstructures, the proposed 2D CNNs based on cross-section images have several advantages. On one hand, the training speed of 2D CNNs can be two orders of magnitude faster than 3D CNNs (0.015 second vs 3.3 second per iteration on a sample), because of the reduced number of parameters and the reduced computational costs. Previous works on 3D CNNs usually consider the input size of ~50×50×50 [41, 42], while in this work, the size of microstructures is increased to 100×100×100, which makes the difference of computational efficiency more significant. Such reduction of computational cost also makes the optimization of the models much easier. On the other hand, 2D cross-section images can be easier to obtain in real experiments than the whole 3D microstructures. Hence the method can offer better opportunity for experimental realization. The drawback of the proposed method is that in order to improve the prediction accuracy, appropriate selection of cross-section images is needed. Therefore, the representativeness of cross-section images is further investigated.

The testing errors decrease with the number or size of cross-section images because of the decrease of uncertainty. Such uncertainty can be estimated by using the volume fraction of the cross-section images as a criterion for representing 3D information, which can also be used to test the appropriateness of the selection of cross-section images. Besides, for anisotropic composites, the representativeness of cross-section images is also related to their directions. Cross-section images along and perpendicular to the preferred directionality of fillers are good choice because they can fully determine the shape and distribution of the fillers. Cross-section images along the heat flow can distinguish serial structures and parallel structures easier than cross-section images perpendicular to the heat flow, which makes them more representative.

*5.2 Summary*

In summary, we use multiple cross-section images as the input of 2D CNN models to predict effective thermal conductivity of 3D composites. The optimized 2D CNN models can achieve similar accuracy with 3D CNN models, and are easily generalized to different types of microstructures, including particle filled composites and stochastic complex composites. The obtained MAE and RMSE are 2.5%, 3.1% and 0.8%, 1.0% for particle filled composites and stochastic complex composites, respectively. When normalized by the standard deviation of the target values, MAE and RMSE are about 7.5% and 10% respectively for both types of microstructures. The proposed method is not restricted to predict effective thermal conductivity of 3D composites, but also applicable to other properties dependent on 3D structures.

**Acknowledgements**

This work is supported by the National Natural Science Foundation of China (No. 51676121). Simulations were performed with computing resources granted by HPC (π) from Shanghai Jiao Tong University. We thank Mostafa Mollaali (Shanghai Jiao Tong University) for his helpful guidance on the fenics package.